\documentclass[useAMS,usenatbib]{mn2e}

\usepackage{amsmath}
\usepackage{amssymb}
\usepackage{bm}
\usepackage{color}
\usepackage{mathrsfs}
\usepackage{graphicx}

\def\vector#1{\mbox{\boldmath $#1$}}

\newcommand{\flag}[1]{\texttt{\lowercase{#1}}}
\newcommand{\galpy}{\flag{galpy}}

\newcommand\araa{{ARA\&A}}%
\newcommand\apj{{ApJ}}%
\newcommand\apjl{{ApJ}}%
\newcommand\apjs{{ApJS}}%
\newcommand\aap{{A\&A}}%
\newcommand\aapr{{A\&A~Rev.}}%
\newcommand\aaps{{A\&AS}}%
\newcommand\mnras{{MNRAS}}%
\newcommand\pasj{{PASJ}}%
\title[Velocity dispersion of the thick disc]{
Vertical kinematics of the thick disc at 4.5 $\lesssim R \lesssim$ 9.5 kpc}
\author[K. Hattori and G. Gilmore]{Kohei Hattori$^{1}$\thanks{E-mail: khattori@ast.cam.ac.uk (KH)} and 
Gerard Gilmore$^{1}$\\
$^{1}$Institute of Astronomy, University of Cambridge, 
Madingley Road, Cambridge CB3 0HA, United Kingdom}
\begin{document}

\date{Accepted 2015 August 12. Received 2015 August 12; in original form 2015 March 23}

\pagerange{\pageref{firstpage}--\pageref{lastpage}} \pubyear{20xx}

\maketitle

\label{firstpage}

\begin{abstract}

We explored a method to reconstruct the distribution function of the Galactic thick disc 
within the action space where nearby thick-disc stars are distributed. 
By applying this method to 
127 chemically-selected thick-disc stars in the Solar neighbourhood, 
we found that 
the vertical velocity dispersion 
that corresponds to the reconstructed distribution function 
declines approximately as $\exp (-R/R_{\rm s})$ 
at $4.5 \;{\rm kpc} \lesssim R \lesssim 9.5 \;{\rm kpc}$, 
with 
$R_{\rm s}=8.3\pm1.1 {\rm (rand.)}\pm1.6 {\rm (sys.)}$ kpc. 
Also, we found that the vertical velocity dispersion $\sigma_z$ of our local thick-disc stars 
shows only weak dependency on radial and azimuthal velocities $(v_R, v_\phi)$. 
We discuss possible implications of these results 
on the global structure of the Milky Way thick disc. 

\end{abstract}


\section{Introduction} \label{section:introduction}

Historically, the thick disc of the Milky Way was first identified through star counts toward the Galactic Poles 
as a vertically extended disc component with scale height $\sim$ 1 kpc \citep{Yoshii1982, Gilmore1983}, 
in addition to the thin disc with scale height $\sim 0.3 \;{\rm kpc}$ 
that dominates the disc stars in the immediate Solar neighbourhood. 
Therefore, at the outset, the tentative definition of the thick-disc stars was 
those disc stars with large vertical orbital excursions and large vertical velocities.

Later, spectroscopic studies (e.g., \citealt{Bensby2003, Bensby2004, Reddy2006}) 
on kinematically-selected disc stars suggested that 
stars with large vertical motions (which are likely to belong to the thick disc) 
tend to show lower [Fe/H] and higher [$\alpha$/Fe] 
than those with small vertical motions (thin-disc stars). 
These chemical properties suggest that 
the thick-disc stars are older than thin-disc stars (lower [Fe/H]) 
and that the star formation timescale of the thick disc 
was shorter than that of the thin disc (higher [$\alpha$/Fe]).

Recently, 
Klaus Fuhrmann \citep{Fuhrmann1998, Fuhrmann2004, Fuhrmann2008, Fuhrmann2011} 
investigated a kinematically-unbiased volume-complete sample of Solar-type disc and halo stars located within 25 pc of the Sun. 
The distribution of his sample stars in the [Fe/H]-[Mg/Fe] space reveals two chemically distinct populations of disc stars (see Figure 15 of \citealt{Fuhrmann2011}). 
Based on the kinematical properties of these populations, 
he identified the lower-[Fe/H] and higher-[Mg/Fe] population to be the thick disc, 
and the other population to be the thin disc. 
This clear separation between the thin and thick discs is also confirmed 
in the nearby (heliocentric distance $d<$ 100 pc) kinematically-unbiased sample of \cite{Adibekyan2012} 
for which [Fe/H] and [$\alpha$/Fe] are available through high-resolution spectroscopy. 
These recent observations suggest 
that the thick disc is better defined by chemical compositions of stars, 
especially by [Fe/H] and [$\alpha$/Fe] 
\citep{Masseron2015}.

In the past decades, 
our understanding of the structure of the thick disc towards the Galactic Poles has been greatly improved \citep{Gilmore1989, Rix2013, Yoshii2013}. 
The next step forward is to unravel its more global structure, 
such as the radial dependence of its vertical structure. 
So far, many attempts have been made to fulfil this goal, 
and they are broadly categorised into two classes of studies.

The first class of studies are based on nearby samples of disc stars. 
One such example is \cite{Binney2012}, 
who fitted the distribution of local disc stars with his distribution function model. 
He discussed some global structure of the thick disc 
by looking into his best-fit models. 
Although this kind of studies can predict almost everything if the best-fit models are reasonably correct, 
one critical problem with these studies is the validity of the functional forms of the assumed distribution functions.

The second class of studies are based on in-situ samples of (relatively) distant disc stars. 
This class can be further categorised into three sub-classes: 
those studies using 
(i) high-resolution spectroscopic samples; 
(ii) medium-resolution spectroscopic samples; or 
(iii) photometric samples. 
The advantage of sub-class (i) studies is that we can define the thick disc purely by means of the stellar chemistry \citep{Bensby2011}. 
However, the number of stars that are currently available is less than a few hundred, 
and this small sample size makes it hard to obtain some statistical properties of distant disc stars. 
Also, since the errors in distance and proper motion are large, 
kinematical analyses are difficult for these stars. 
In the sub-class (ii) studies, 
much larger sample of stars are available than in the sub-class (i) studies. 
A recent example of this sub-class is \cite{Bovy2012a}, 
who studied the density distribution of chemically defined disc populations 
by fitting the SDSS/SEGUE data with analytic models of the density profile. 
However, since the chemical abundances of these samples are less accurate than high-resolution samples, 
some blending of the thin- and thick-disc stars is inevitable in this kind of studies.  
Most of the sub-class (iii) studies are based on the star count method (e.g., \citealt{Robin2014}). 
Photometric samples 
are advantageous in grasping the global structure of the stellar disc 
because the available sample size is the largest among these subclasses, 
and because the sample stars cover a wide range of heliocentric distance. 
However, since the photometric data lack chemical information for each sample star, 
it is not possible to separate the sample into stellar components. 
This inseparability means that 
one needs to adequately model all the stellar components that contribute to the sample, 
which is harder than adequately modelling the thick disc only.

By taking into account these problems, 
in this paper we explore 
the possibility of 
constraining the global structure of the thick disc 
based on a small number of nearby chemically-selected stars, 
but not relying on analytic modelling of the thick disc. 
Specifically, 
we reconstruct the distribution function of the thick disc 
within a certain portion of the action space 
which is accessible from local observations.

This paper is organised in the following manner. 
First, we describe our sample stars in section \ref{section:sample}. 
In section \ref{section:theory}, 
we introduce the concepts of the observable action space $\Gamma_\odot$ 
and the observable distribution function $f_\odot$. 
There, we discuss the possibility of inferring 
some information on the velocity dispersion of the thick disc by using $f_\odot$. 
In section \ref{section:method}, we present our method to reconstruct $f_\odot$ from a local sample of thick-disc stars. 
In section \ref{section:analysis}, we present our main results. 
The implications from our study are discussed in section \ref{section:discussion}, 
and section \ref{section:conclusion} sums up our analyses.

\section{Sample} \label{section:sample}

\subsection{The coordinate system and some numerical constants}

In this paper, 
we assume that the Milky Way potential is axisymmetric 
and adopt the usual Galactocentric cylindrical coordinate system $(R, \phi, z)$ to describe 
the three-dimensional positions of stars in the Milky Way. 
The three-dimensional velocities along the respective coordinates are denoted by 
$(v_R, v_\phi, v_z)$, where $v_R$ is positive outwards, $v_\phi$ is positive in the direction of Galactic rotation, 
and $v_z$ is positive in the direction of the North Galactic Pole. 

We assume that the local standard of rest (LSR)  
is in a circular orbit 
with a rotation speed of 
$v_{\rm LSR} = 240 \;{\rm km\;s^{-1}}$ 
\citep{Reid2014}
and that the Galactocentric distance of the Sun is 
$R_0 = 8.3 \;{\rm kpc}$. 
We also adopt the peculiar motion of the Sun with respect to the LSR 
determined by \cite{Schonrich2010}. 

\subsection{Construction of the local thick-disc sample}

In this paper, 
we assume that the thick disc of the Milky Way 
is a single stellar population and 
define the thick-disc stars to be those stars 
with $-1.0 \leq {\rm [Fe/H]} \leq -0.2$ and ${\rm [Mg/Fe]} \geq 0.2$, 
by taking into account the Solar neighbour observations of disc stars (e.g., \citealt{Fuhrmann2011}). 
We constructed a kinematically-unbiased sample of 127 thick-disc stars,  
for which accurate position, velocity, [Fe/H], and [Mg/Fe] are available. 
84 stars in our sample were taken from \cite{Adibekyan2012}, 
and the other 43 stars were taken from \cite{Soubiran2005}. 
Most of our sample stars are distributed within 100 pc from the Sun.

\subsubsection{Sample from Adibekyan et al. (2012)}

\cite{Adibekyan2012} published a kinematically-unbiased sample of 1111 stars, 
for which accurate [Fe/H] and [Mg/H] estimated from high-resolution spectroscopic observations
as well as the three dimensional Cartesian velocity based on Hipparcos observations 
are available. 
We derived the distances and the line-of-sight velocities for 1103 stars in their sample 
that were also included in the updated Hipparcos catalog of \cite{vL2007}. 
In the course of this inversion,
we took into account the Solar peculiar motion adopted in \cite{Adibekyan2012} 
and we used the three-dimensional Cartesian velocity taken from \cite{Adibekyan2012} 
and the position on the sky and the proper motion taken from \cite{vL2007}.  
We then calculated the three-dimensional velocity $(v_R, v_\phi, v_z)$ in our coordinate system, 
by using our assumption on $(R_0, v_{\rm LSR})$. 
From these 1103 stars, 
we selected those stars 
that satisfy both the thick-disc chemical criteria ($-1.0 \leq {\rm [Fe/H]} \leq -0.2$ and ${\rm [Mg/Fe]} \geq 0.2$) 
and the 
kinematical criteria of 
$v_\phi \geq 70\;{\rm km\;s^{-1}}$ 
and $\sqrt{v_R^2 + v_z^2} \leq 160\;{\rm km\;s^{-1}}$. 
The latter velocity cuts were applied to minimise possible contamination from halo stars. 
Finally, we discarded those stars 
that had been flagged as non-single stars, 
and we obtained 84 stars.

\subsubsection{Sample from Soubiran \& Girard (2005)}
\cite{Soubiran2005} published a catalog of 743 stars compiled from 11 catalogs of nearby stars 
and listed the chemical abundances of these stars.\footnote{
When more than one published data are available for a certain chemical abundance of a given star, they list the averaged value.} 
Among these 11 catalogs, those catalogs taken from 
\cite{AP2004}, \cite{Chen2000}, \cite{Edvardsson1993}, \cite{Mishenina2004}, and \cite{Reddy2003} are kinematically unbiased, 
so we first selected those stars that are included in either of these five catalogs. 
Then, we excluded those stars that happened to be included in \cite{Adibekyan2012}. 
The position and velocity of these stars were then calculated either 
by using the data from Genova-Copenhagen Survey \citep{Holmberg2009} 
or by using the updated Hipparcos catalog \citep{vL2007} and SIMBAD database. 
Finally, as in the previous subsection, 
we applied the chemical and velocity criteria and obtained 43 thick-disc stars.

\subsubsection{Caveats on the velocity cuts}
As described above, we adopted a pair of velocity cuts to exclude halo stars. 
Rigorously speaking, these cuts potentially could have introduced some kinematical bias on our analyses. 
However, 
we have confirmed that 
adopting a looser velocity constraint of $v_\phi>0\;{\rm km\;s^{-1}}$ (without cuts on $v_R$ or $v_z$) 
has only negligible effects on our final results.

\subsection{Chemical and Kinematical properties of our sample}

\begin{figure}
\begin{center}
	\includegraphics[angle=0,width=0.95\columnwidth]{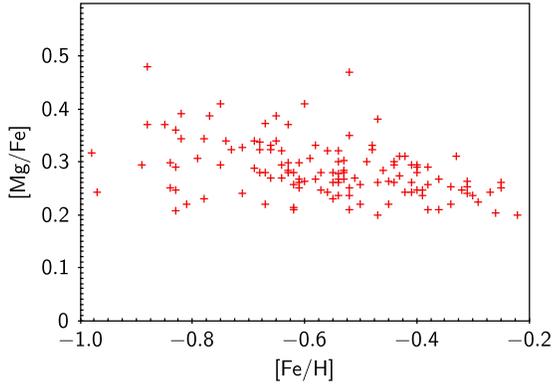} \\
\end{center}
\caption{
Chemical abundances of our thick-disc sample. 
}
\label{fig:FeH_MgFe}
\end{figure}

The distribution of our sample stars in the [Fe/H]-[Mg/Fe] plane is shown in Figure \ref{fig:FeH_MgFe}. 
The uncertainties in the chemical abundances are small enough 
(typical uncertainties are $\Delta{\rm[Fe/H]} =$ 0.03 dex and $\Delta{\rm [Mg/Fe]} = $ 0.05 dex for the \citealt{Adibekyan2012} sample) 
compared with the scatter in this diagram. 
Therefore, our sample does not suffer from severe contamination from low-[Mg/Fe] stars.  

\begin{figure*}
\begin{center}
	\includegraphics[angle=0,width=0.75\columnwidth]{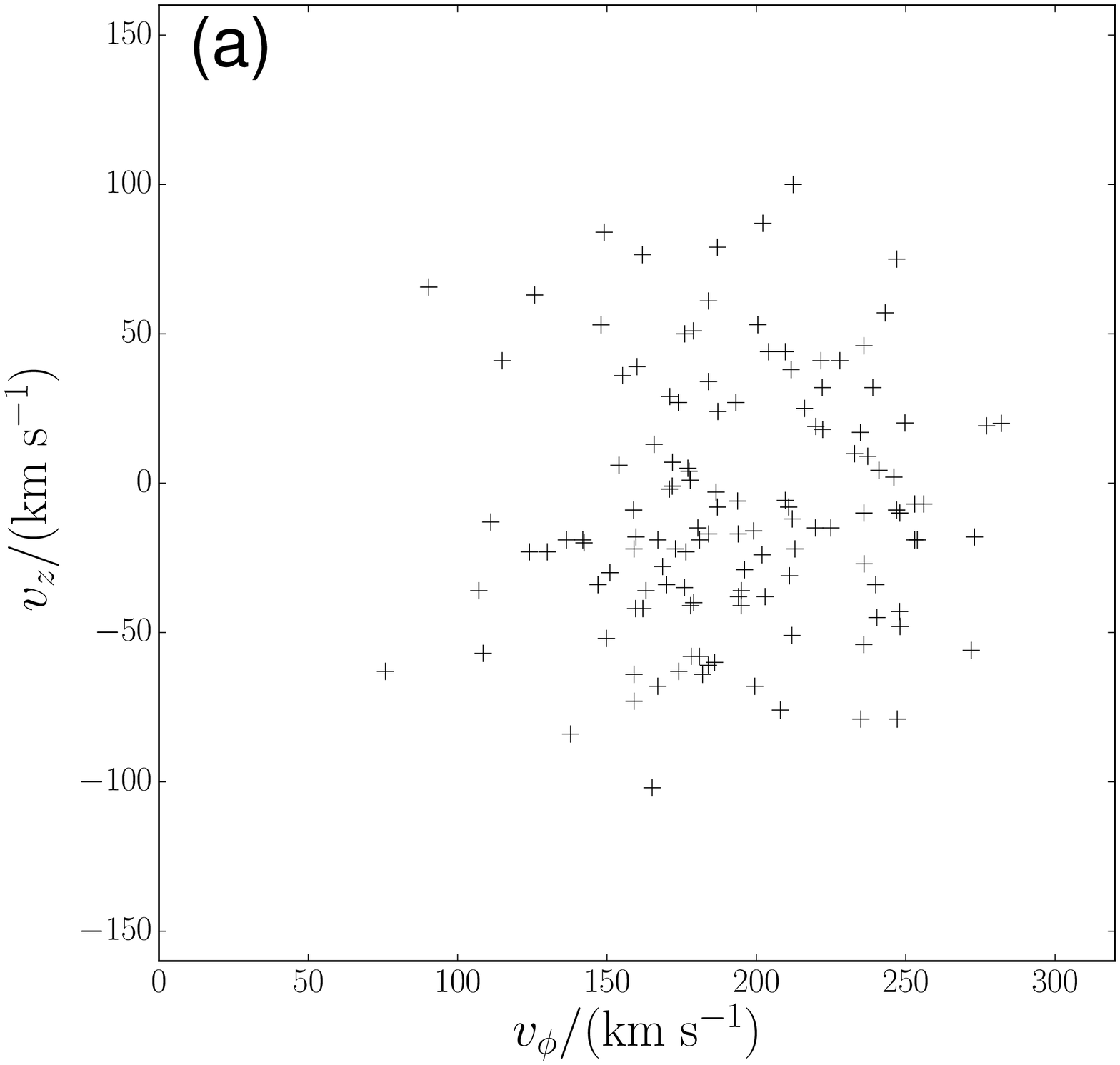} 
	\includegraphics[angle=0,width=0.75\columnwidth]{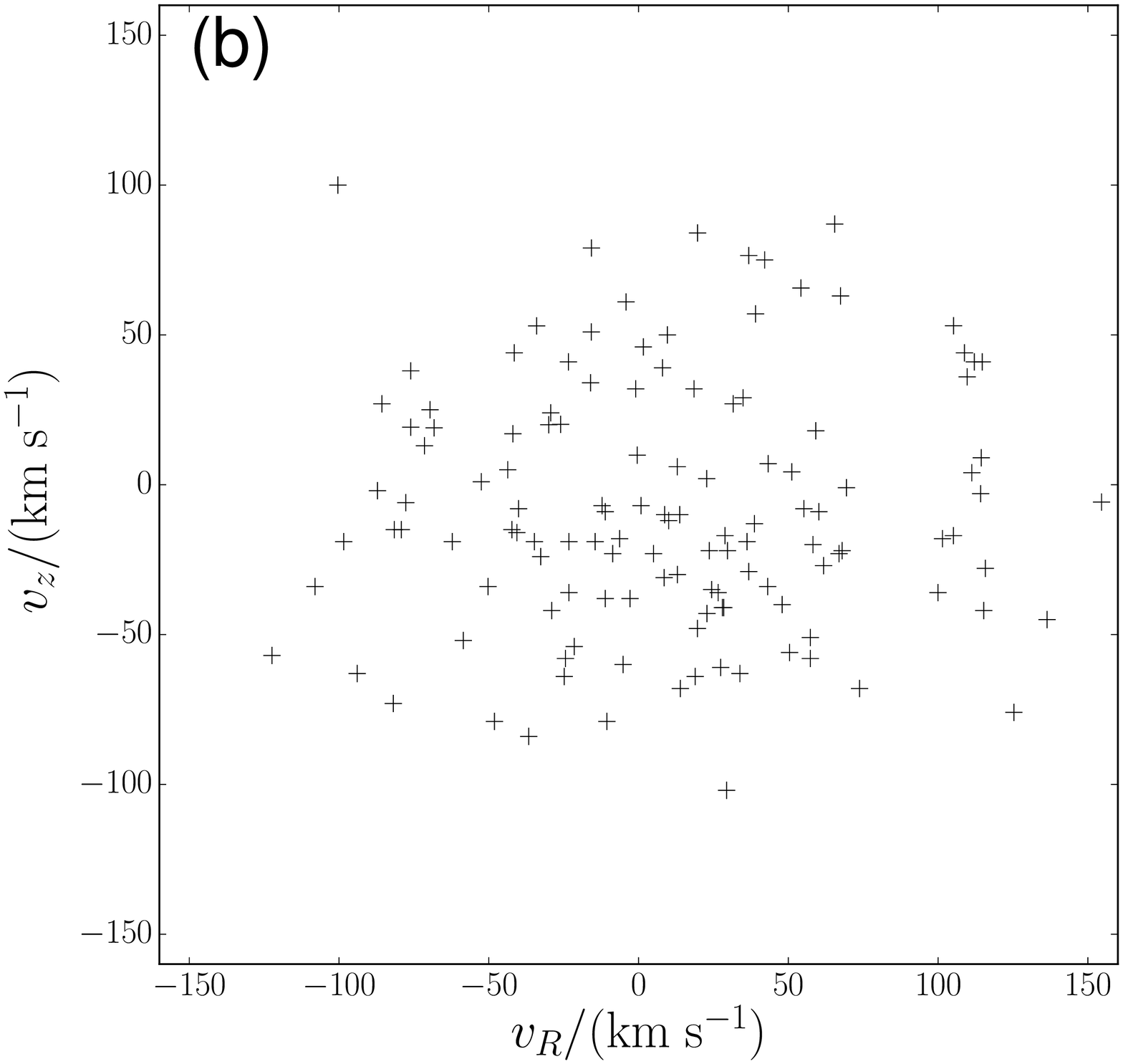} \\
	\includegraphics[angle=0,width=0.75\columnwidth]{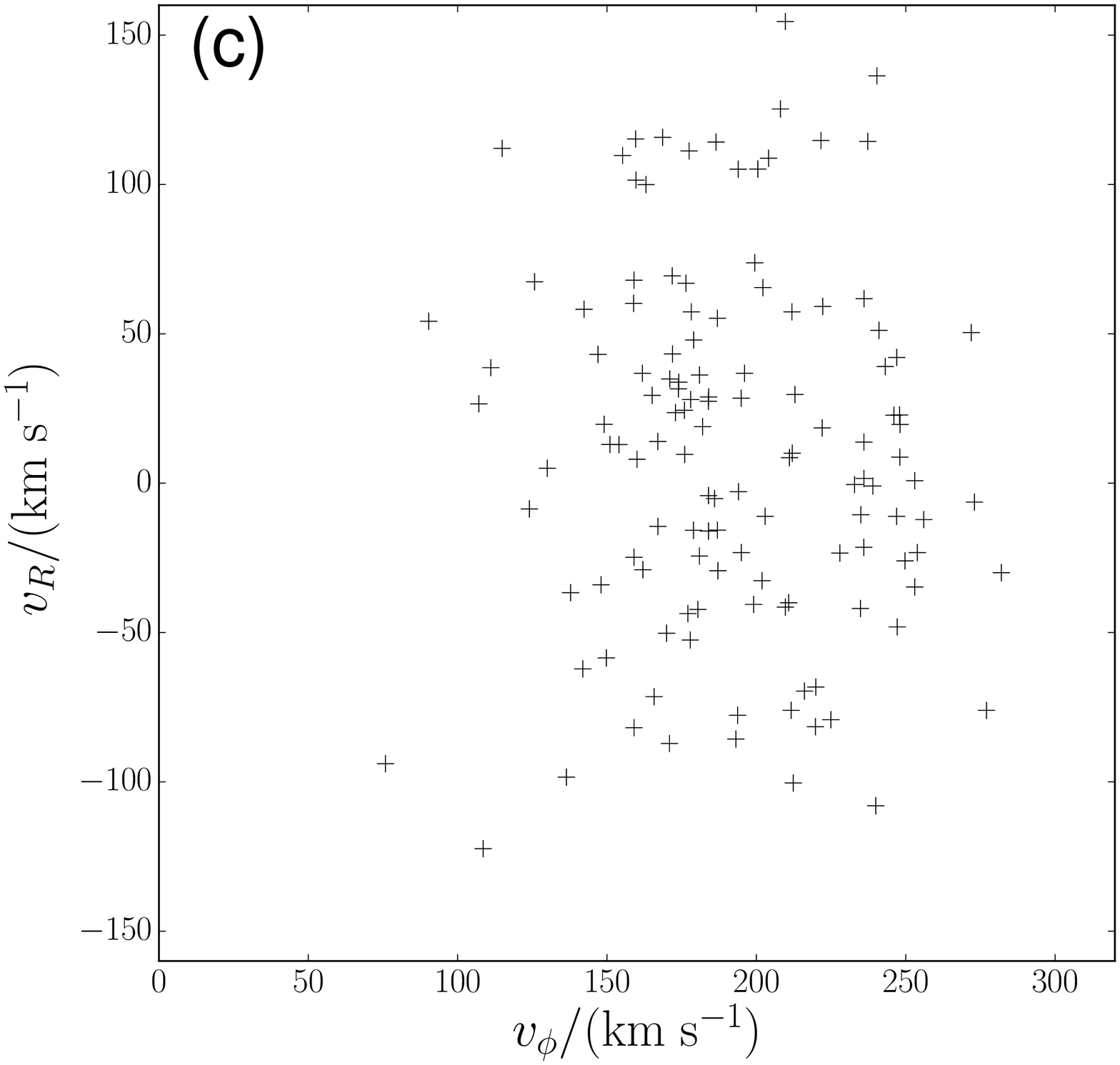} 
	\includegraphics[angle=0,width=0.75\columnwidth]{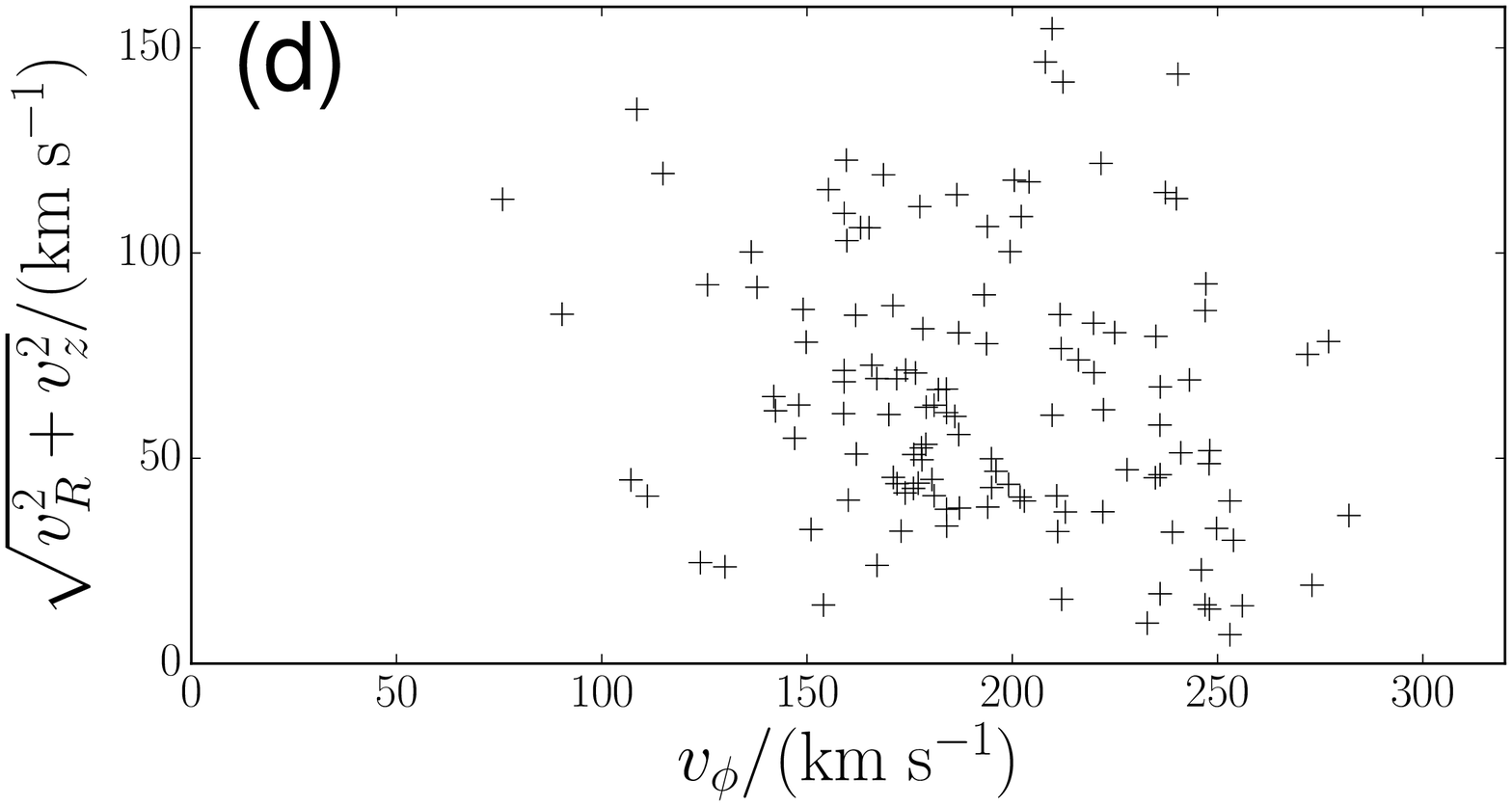} \\
\end{center}
\caption{
Velocity distributions of our thick-disc sample in 
(a) $(v_\phi, v_z)$-space, 
(b) $(v_R, v_z)$-space, 
(c) $(v_\phi, v_R)$-space, 
and 
(d) $\left( v_\phi, \sqrt{v_R^2+v_z^2} \right)$-space. 
}
\label{fig:velocity_distribution}
\end{figure*}

Figure \ref{fig:velocity_distribution} shows 
the velocity distribution of our sample in the $(v_\phi, v_R)$, $(v_\phi, v_z)$, $(v_R, v_z)$ and $(v_\phi, \sqrt{v_R^2+v_z^2})$ planes. 
Our sample stars show 
a mean velocity of 
$(\langle v_R \rangle, \langle v_\phi \rangle, \langle v_z \rangle) = (10.5\pm5.3, 192.3\pm3.6, -7.9\pm3.7) \;{\rm km\;s^{-1}}$, 
and thus lag from the LSR by nearly $50 \; {\rm km\;s^{-1}}$. 
The three-dimensional velocity dispersion is 
$(\sigma_R, \sigma_\phi, \sigma_z) = (59.6, 40.9, 41.6)\;{\rm km\;s^{-1}}$.

\begin{figure}
\begin{center}
	\includegraphics[angle=-90,width=0.9\columnwidth]{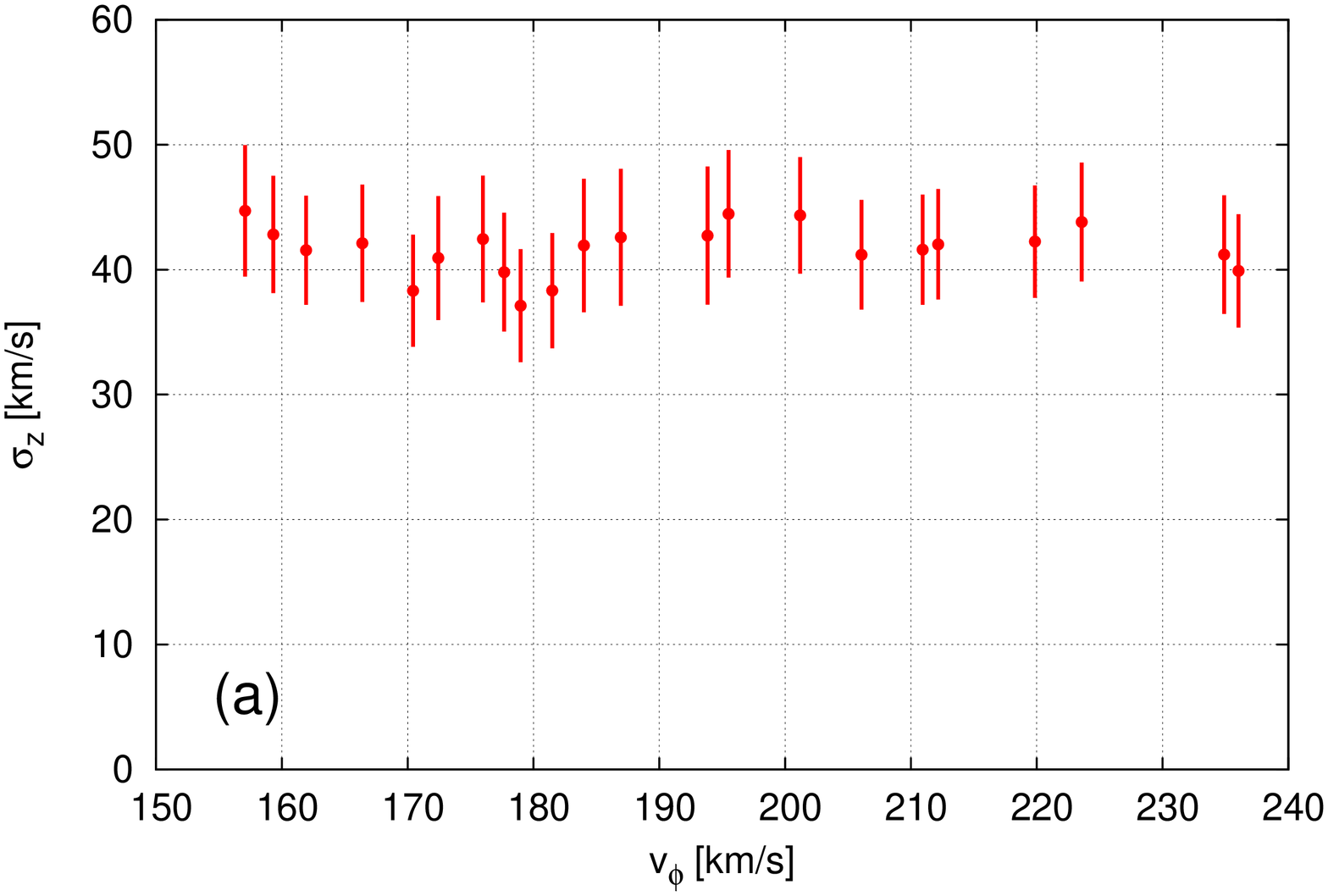} \\
	\includegraphics[angle=-90,width=0.9\columnwidth]{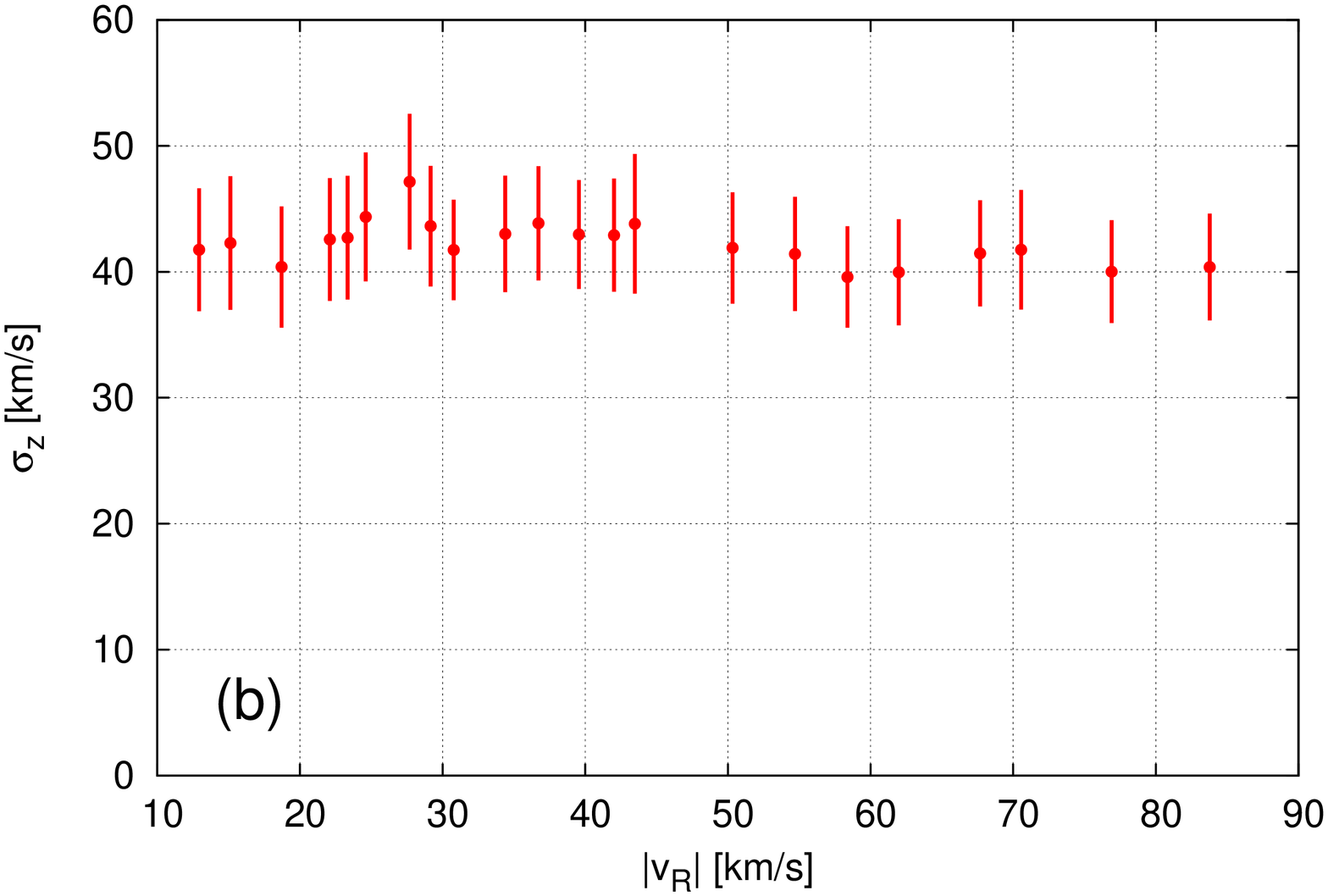} \\
\end{center}
\caption{
Vertical velocity dispersion $\sigma_z$ of our thick-disc sample 
as a function of 
(a) $v_\phi$ and (b) $|v_R|$. 
Each data point represents 
the value of $\sigma_z$ of a binned sample of 40 stars, 
and each pair of adjacent bins share 36 stars in common. 
The uncertainty in $\sigma_z$ is calculated 
by assuming Gaussian distribution of vertical velocity $v_z$ in each bin. 
}
\label{fig:separability}
\end{figure}

In Figure \ref{fig:separability}, we show the value of $\sigma_z$ 
as functions of 
$v_\phi$ and $|v_R|$. 
To make these plots, 
we first 
sort the sample in 
$v_\phi$ or $|v_R|$, 
evaluate the dispersion $\sigma_z$ for a binned sample of $N=40$ stars, 
and move through the sample in steps of 4 stars (so that any pair of adjacent bins share 36 stars). 
The derived value of $\sigma_z$ for each bin is 
plotted against the median value of 
$v_\phi$ or $|v_R|$ 
of the bin.

Although the bin size of 40 stars is rather large compared with the total sample size of 127 stars, 
the result of this figure suggests that 
there is no strong trend in $\sigma_z$ 
as a function of 
$|v_R|$ or $v_\phi$. 
This property suggests that the vertical kinematics ($\sigma_z$) 
is not strongly correlated with the in-plane motion ($|v_R|$ or $v_\phi$).

\section{Theory}\label{section:theory}

\subsection{Observable distribution function}

Most of the thick-disc stars in the Solar neighbourhood are on non-circular orbits. 
More specifically, 
the majority of nearby thick-disc stars have guiding radii smaller than the Galactocentric distance of the Sun ($R_0$), 
due to the asymmetric drift \citep{BT2008}. 
Therefore, if the thick disc is a static stellar system, 
kinematics of nearby thick-disc stars carry 
some information on the thick disc outside the Solar neighbourhood, 
especially on the inner part of the thick disc.

Of course, 
nearby thick-disc stars do not carry all the information of the thick disc, 
since these stars are inherently biased in that 
they represent a portion of the thick-disc stars whose orbits penetrate through the Solar neighbourhood. 
In other words, a nearby sample only covers a certain fraction of action space $\Gamma_\odot$ 
where the corresponding orbits bring stars to the Solar neighbourhood \citep{May1986}. 
Hence, the best we can do with nearby samples 
is to reconstruct the distribution function $f({\vector J})$ within this {\it observable action space} $\Gamma_\odot$, 
i.e. to reconstruct the {\it observable distribution function} defined by
\begin{equation} \label{eq:fobs}
	f_\odot ({\vector J}) = 
	\begin{cases}
		f({\vector J}) & ({\vector J} \in \Gamma_\odot)\\
		0 & ({\rm otherwise}). 
	\end{cases} 
\end{equation}
We here note that $f_\odot ({\vector J})$ is different from the as-observed action distribution. 
The fraction of its orbital period that a given star spends in the Solar neighbourhood depends on its orbit, 
so that the probability of finding the star in the Solar neighbourhood depends on its action. 
In section \ref{section:method}, 
we take into account this probability 
to reconstruct the observable distribution function $f_\odot ({\vector J})$.

\subsection{Observable vertical velocity dispersion}\label{subsection:observable_dispersion}

\begin{figure}
\begin{center}
	\includegraphics[angle=0,width=0.9\columnwidth]{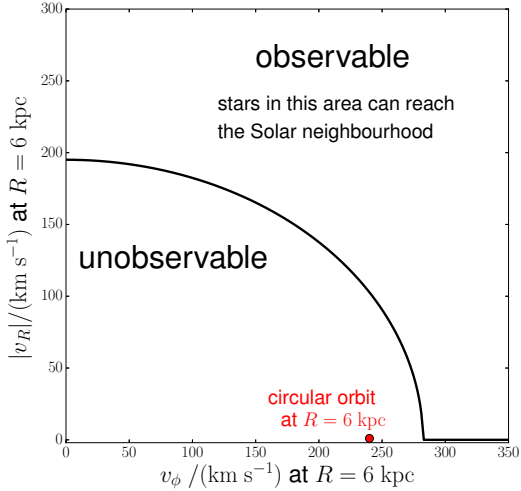} 
\end{center}
\caption{
The allowed region in the $(v_R, v_\phi)$-space 
for disc stars at $(R, z)=(6,0)$ kpc with $v_z=0\;{\rm km\;s^{-1}}$ 
to reach the Solar neighbourhood, 
calculated for the potential model in \ref{section:analysis}. 
The boundary curve between 
the observable region and unobservable region 
corresponds to $R_{\rm apo}=R_0=8.3\;{\rm kpc}$ 
($R_{\rm apo}$ is the apocentric distance of the orbit), 
and thus the circular orbit at $R=6$ kpc (red dot)
lies within the unobservable region. 
When $|v_z|>0\;{\rm km\;s^{-1}}$, the boundary curve 
slightly shifts downwards in this diagram. 
However, this difference is not significant 
for a typical value of $|v_z|$ of the thick-disc stars.  
In this plot, 
we assume a realistic potential model of the Milky Way 
described in section \ref{section:analysis}. 
}
\label{fig:observable}
\end{figure}

In order to have a physical insight into $\Gamma_\odot$, 
let us consider the velocity space that corresponds to $\Gamma_\odot$ 
by assuming a potential model of the Milky Way. 
Figure \ref{fig:observable} 
shows the allowed range of $(v_R, v_\phi)$ 
for disc stars located at $(R,z)=(6, 0)$ kpc with zero vertical velocity 
to reach the Solar neighbourhood. 
Since only a portion of the disc stars at $R=6$ kpc 
can reach the Solar neighbourhood, 
what we can learn from $f_\odot$ 
about the thick disc at $R=6$ kpc  
is the velocity distribution of stars 
within the `observable' region in Figure \ref{fig:observable}.

With the information on $f_\odot$, 
we can estimate the {\it observable vertical velocity dispersion} defined by 
\begin{align}
&\sigma_{z, \odot}^2 (R, 0) \equiv 
\frac{\int d^3{\vector v} \; v_z^2 f_\odot({\vector J})}{\int d^3{\vector v} \; f_\odot({\vector J})} 
\label{eq:def_sigz_fobs} 
\end{align} 
at any given radius $R$ on the disc plane. 
This dispersion corresponds to the 
vertical velocity dispersion of 
those stars located at $R$ 
whose velocities are within the observable region. 
Thus $\sigma_{z,\odot}$ is conceptually different from 
the (ordinary) vertical velocity dispersion given by
\begin{align}
&\sigma_z^2 (R, 0) = \frac{\int d^3{\vector v} \; v_z^2 f({\vector J})}{\int d^3{\vector v} \; f({\vector J})}, 
\label{eq:def_sigz_f} 
\end{align} 
which is the vertical velocity dispersion of all the stars located at $R$.

\section{Method} \label{section:method}

Here we describe 
how to reconstruct $f_\odot ({\vector J})$ 
of the thick disc from a nearby sample of thick-disc stars, 
based on the idea of \cite{SLZ1990}. 
We note that \cite{McMillan2012} investigated 
a rigorous treatment of an as-observed distribution function 
in which survey selection function is taken into account (see Appendix). 
We do not follow their formulation here 
since we do not know the exact survey selection function for our sample stars 
(but we do know that our sample stars are not kinematically selected). 
Although we do not take into account the survey selection function, 
our approach has an advantage that 
we do not have to assume the functional form of the distribution function.

\subsection{Reconstruction of observable distribution function $f_\odot$ of the thick disc} 
\label{section:method_reconstruction}

In the following, 
we regard disc stars as test particles moving in the Milky Way potential. 
Also, we assume that we have a kinematically-unbiased sample of 
$N$ thick-disc stars which happen to be passing near the Sun.

Under an axisymmetric potential $\Phi(R,z)$, 
each star has three actions ${\vector J} = (J_R, J_\phi, J_z)$ \citep{BT2008, Binney2010}. 
Let us denote the actions of the $i$th star as ${\vector J}_i$. 
Then, the time-averaged distribution function of a single orbit characterised by ${\vector J}_i$ 
is expressed as 
\begin{align}
f_i^{\rm single} ({\vector x}, {\vector v}) &=
\frac{1}{{(2\pi)}^3} \delta ({\vector J}({\vector x}, {\vector v}) - {\vector J}_i) ,
\end{align}
and the corresponding spatial density is given by 
\begin{align} \label{eq:single_orbit_density}
\rho({\vector J}_i; {\vector x}) = \int d^3 {\vector v} \; f_i^{\rm single} 
= \frac{1}{{(2\pi)}^3} {\left| \frac{\partial {\vector v}}{\partial {\vector J}} \right|}_{\left({\vector J}_i, {\vector x}\right)}. 
\end{align}
The value of $\rho({\vector J}_i; {\vector x})$ can be interpreted as 
the probability density of finding a star with action ${\vector J}_i$ at position ${\vector x}$, 
since $\int d^3 {\vector x}d^3 {\vector v} \; f_i^{\rm single} = 1$ \citep{CB2001}.

By following  \cite{SLZ1990},
we approximate the observable distribution function $f_\odot$ 
to be the weighted sum of single orbit distribution functions given by 
\begin{align} \label{eq:fobs_discrete}
f_\odot ({\vector J}) = \sum_{i=1}^N s_i w_i f_i^{\rm single} ({\vector x}, {\vector v}) . 
\end{align}
Here, $s_i$ is a selection index that determines 
whether the $i$th star should be included ($s_i=1$) or not ($s_i=0$), 
which shall be introduced to stabilise our analyses. 
The factor $w_i(>0)$ is the orbit weighting factor to be determined. 
Although \cite{SLZ1990} evaluated $w_i$ through Bayesian statistics, 
here we numerically evaluate $\rho({\vector J}_i; {\vector x}_i)$ 
and simply set $w_i = 1/ \rho({\vector J}_i; {\vector x}_i)$. 
This simple choice of $w_i$ corresponds to 
a limiting case where 
all of our sample stars have identical position \citep{SLZ1990}.\footnote{
This property can be intuitively understood in the following manner.  
The probability $p$ of finding a given star in the Solar neighbourhood 
is equal to the fraction of its orbital period to stay in the Solar neighbourhood. 
Thus, if such a star is observed in the Solar neighbourhood, 
it represents $1/p$ stars in the same orbit (including those with different orbital phases). 
}
Since most of our sample stars are located 
within 100 pc away from the Sun, 
our simple choice of $w_i$ is well justified.

\subsection {Reconstruction of vertical velocity dispersion of the thick disc}  \label{subsection:method_sigz}

Once $f_\odot$ is reconstructed, 
we can estimate the observable vertical velocity dispersion $\sigma_{z,\odot}$ 
at a given point on the disc plane, 
by using equation (\ref{eq:def_sigz_fobs}). 
It can be shown that $\sigma_{z,\odot}(R_0, 0) = \sigma_{z}(R_0, 0)$ 
if the assumed potential is correct and a large enough number of local sample stars are available. 
Thus, we can perform a sanity check of the adopted potential 
by seeing if these dispersions reasonably agree with each other.

\section{Analysis and Results} \label{section:analysis}

\subsection {Analysis: Derivation of weighting factors}
\label{subsubsection:analysis_realistic}

We adopt 
a realistic potential model 
of the Milky Way, 
consisting of halo, bulge, and disc components.

In our model, 
the halo potential  
is given by the Navarro-Frenk-White potential 
\citep{Navarro1996}
of the form 
\begin{equation}
\Phi_{\rm halo} = - v_{\rm H}^2 \frac{\ln \left( 1 + r/d_{\rm H} \right)}{r/d_{\rm H}} , 
\end{equation}
where $r$  
is the Galactocentric distance, 
$v_{\rm H}=379.79\;{\rm km\;s^{-1}}$, 
and 
$d_{\rm H}=9.1\;{\rm kpc}$. 
The bulge potential 
is given by  
\begin{equation}
\Phi_{\rm bulge} = - \frac{GM_{\rm B}}{c_{\rm B} + r}  
\end{equation}
\citep{Hernquist1990}. 
We adopt $M_{\rm B} = 2.05\times 10^{10} M_\odot$ and 
$c_{\rm B} = 2\; {\rm kpc}$. 
We have confirmed that adopting a slightly flattened bulge potential 
hardly affects our results. 
For the disc potential, 
we assume the Miyamoto-Nagai potential \citep{Miyamoto1975} given by 
\begin{equation}
\Phi_{\rm discMN} = - \frac{GM_{\rm D}}{ \sqrt{R^2 + {\left( R_{\rm D} + \sqrt{z^2 + z_{\rm D}^2} \right)}^2 } }  
\end{equation} 
where 
$M_{\rm D} = 5.77 \times 10^{10} M_\odot$, 
$R_{\rm D} = 3.2 \;{\rm kpc}$, 
and 
$z_{\rm D} = 0.23 \;{\rm kpc}$. 
The values of parameters are chosen 
by using the publicly available\footnote{Available at http://github.com/jobovy/galpy.} 
code of \galpy\ \citep{Bovy2015}, 
while fixing $R_0 = 8.3 \;{\rm kpc}$, $v_{\rm LSR}=240\;{\rm km\;s^{-1}}$, 
and $c_{\rm B} = 2\; {\rm kpc}$.

After setting up the potential models, 
we derived the weighting factor for the $i$th star, $w_i$, in the following manner. 
We first integrated the orbit of the $i$th star forward in time for a long enough period. 
We then calculated the fraction $\delta f$ of the integration time 
which the star spent in a volume defined by 
$R_i-\delta R < R < R_i + \delta R$, $z_i -\delta z < z < z_i + \delta z$, and $0 \leq \phi < 2 \pi$. 
Here, we denote the observed position of the $i$th star on the meridional plane as $(R_i, z_i)$ 
and we adopted $(\delta R, \delta z) = (0.05, 0.1)\;{\rm kpc}$. 
By taking into account the relation $\delta f = \rho ({\vector I}_i; {\vector x}_i) \times 8 \pi R_i \delta R \delta z$, 
we derived $\rho ({\vector I}_i; {\vector x}_i)$ and set $w_i = 1 / \rho ({\vector I}_i; {\vector x}_i)$. 
We set $s_i=0$ for those stars that satisfy $w_i > \langle \{w_j\} \rangle + 3 \sigma_{\{w_j\}}$, 
where $\langle \{w_j\} \rangle$ and $\sigma_{\{w_j\}}$ are the mean and the standard deviation of $\{ w_j \}$, respectively.  
This final procedure was included to prevent our results being affected by small number of stars with large $w_i$.

\subsection{Reconstructed vertical velocity dispersion}\label{subsection:result_sigma_z}

\begin{figure}
\begin{center}
	\includegraphics[angle=-90,width=0.95\columnwidth]{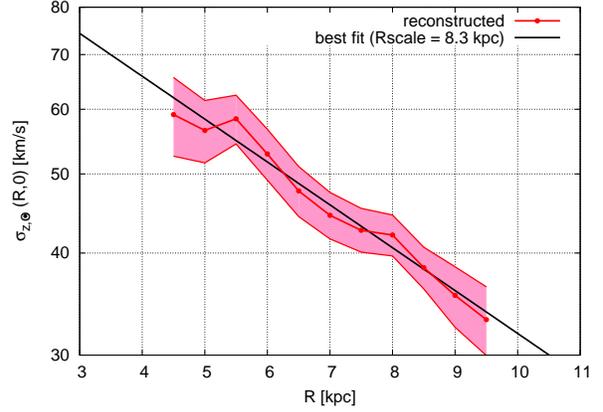}\\
\end{center}
\caption{
The reconstructed observable vertical velocity dispersion $\sigma_{z,\odot}$ 
of the thick disc as a function of $R$. 
}
\label{fig:sigz_data}
\end{figure}

Figure  \ref{fig:sigz_data}  shows the reconstructed profile of $\sigma_{z,\odot}$. 
Here we use an approximation of 
\begin{align} \label{eq:approximate_sigz_realistic}
\sigma_{z,\odot}^2 (R, 0) 
\simeq 
\frac{\sum_i s_i w_i \int_{0}^{\Delta z} dz  \int_{R-\Delta R}^{R+\Delta R} dR^{\prime} \; 2\pi R^{\prime} \int d^3{\vector v} \; v_z^2  f_i^{\rm single}}
{\sum_i s_i w_i \int_{0}^{\Delta z} dz  \int_{R-\Delta R}^{R+\Delta R} dR^{\prime} \; 2\pi R^{\prime} \int d^3{\vector v} \;  f_i^{\rm single}} . 
\end{align} 
We adopt $(\Delta R, \Delta z) = (0.25, 0.1) \;{\rm kpc}$ 
and calculate $\sigma_{z,\odot} (R, 0)$ at 
$4.5 \leq R/{\rm kpc} \leq 9.5$ 
in steps of $0.5\; {\rm kpc}$. 
The radial range of $R$ is set so that more than 42 stars (i.e., more than $N/3$ stars) 
have non-zero contribution to the integrals in equation 
(\ref{eq:approximate_sigz_realistic}), 
i,e., have positive values of $\left(s_i \int_{R-\Delta R}^{R+\Delta R} dR^{\prime} \; 2\pi R^{\prime} \int d^3{\vector v} \; f_i^{\rm single} \right)$.

In Figure \ref{fig:sigz_data}, 
the red dots show the reconstructed values of  $\sigma_{z,\odot}(R, 0)$, 
and the shaded region represents the uncertainty which is derived from a bootstrap resampling of the contributing stars. 
The velocity dispersion declines nearly exponentially as a function of $R$, 
and the best fit exponential profile of $\sigma_{z,\odot}$ is shown with a black straight line. 
We find that 
the best-fit scale length defined by  
\begin{align} \label{eq:Rs}
R_{\rm s} \equiv - \left( \frac{d\; \ln \sigma_{z,\odot} (R, 0)}{d\; R}\right)^{-1} 
\end{align}
is 
$R_{\rm s} = 8.3 \pm 1.1 \mathrm{(rand.)} \pm 1.6 \mathrm{(sys.)} \;{\rm kpc}$. 
The random error is the uncertainty in the fit and 
the systematic error is estimated by performing mock catalog analyses.

We interpolate the values of $\sigma_{z,\odot}$ at $R=8.0$ kpc and $R=8.5$ kpc 
to obtain 
$\sigma_{z, \odot} (R_0, 0) = 39.4 \pm 2.5  \;{\rm km\;s^{-1}}$. 
This value is in agreement with the local value of $\sigma_z = 41.6 \;{\rm km\;s^{-1}}$ in our sample, 
indicating that the adopted potential is reasonable at least in the Solar neighbourhood.

\subsubsection{Dependency on the adopted potential parameters}\label{subsection:result_potential_dependece}

We vary the parameters $(R_{\rm D}, z_{\rm D})$ 
within the range of $2<R_{\rm D}/{\rm kpc}<4$ and $0.15<z_{\rm D}/{\rm kpc}<0.5$ 
and find that 
the nearly exponential profile of $\sigma_{z,\odot}$ is 
a generic feature independent of the assumed parameters. 
$R_{\rm s}$ mildly depends on these parameters, 
but it only increases from 7.5 kpc to 9.4 kpc 
when we vary $R_{\rm D}$ from 2 kpc to 4 kpc; 
and from 8.1 kpc to 9.1 kpc 
when we vary $z_{\rm D}$ from 0.15 kpc to 0.5 kpc.


\subsection{Dependence of the vertical kinematics on $v_R$ and $v_\phi$}

\begin{figure}
\begin{center}
	\includegraphics[angle=0,width=0.95\columnwidth]{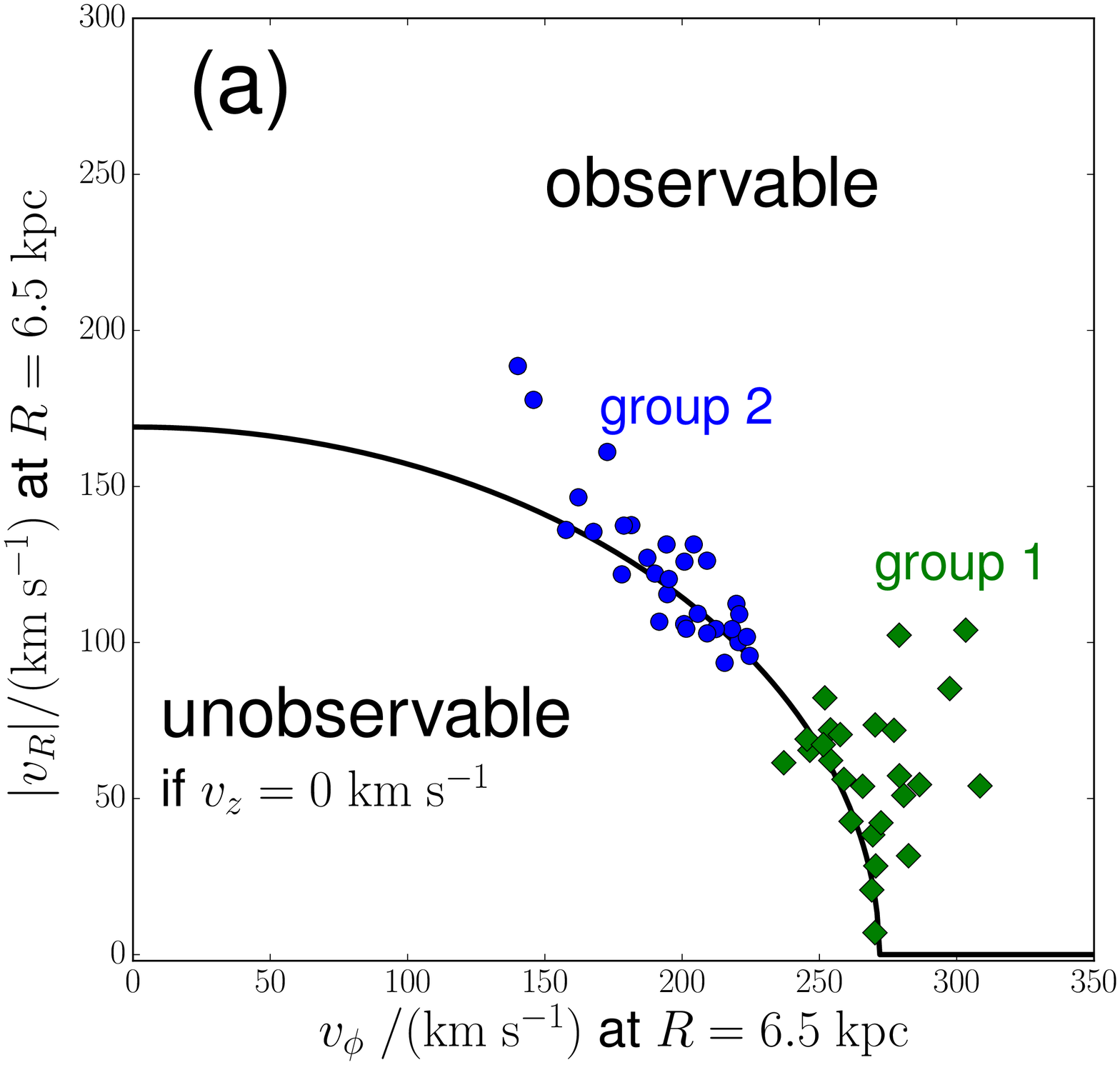} \\
	\includegraphics[angle=-90,width=0.95\columnwidth]{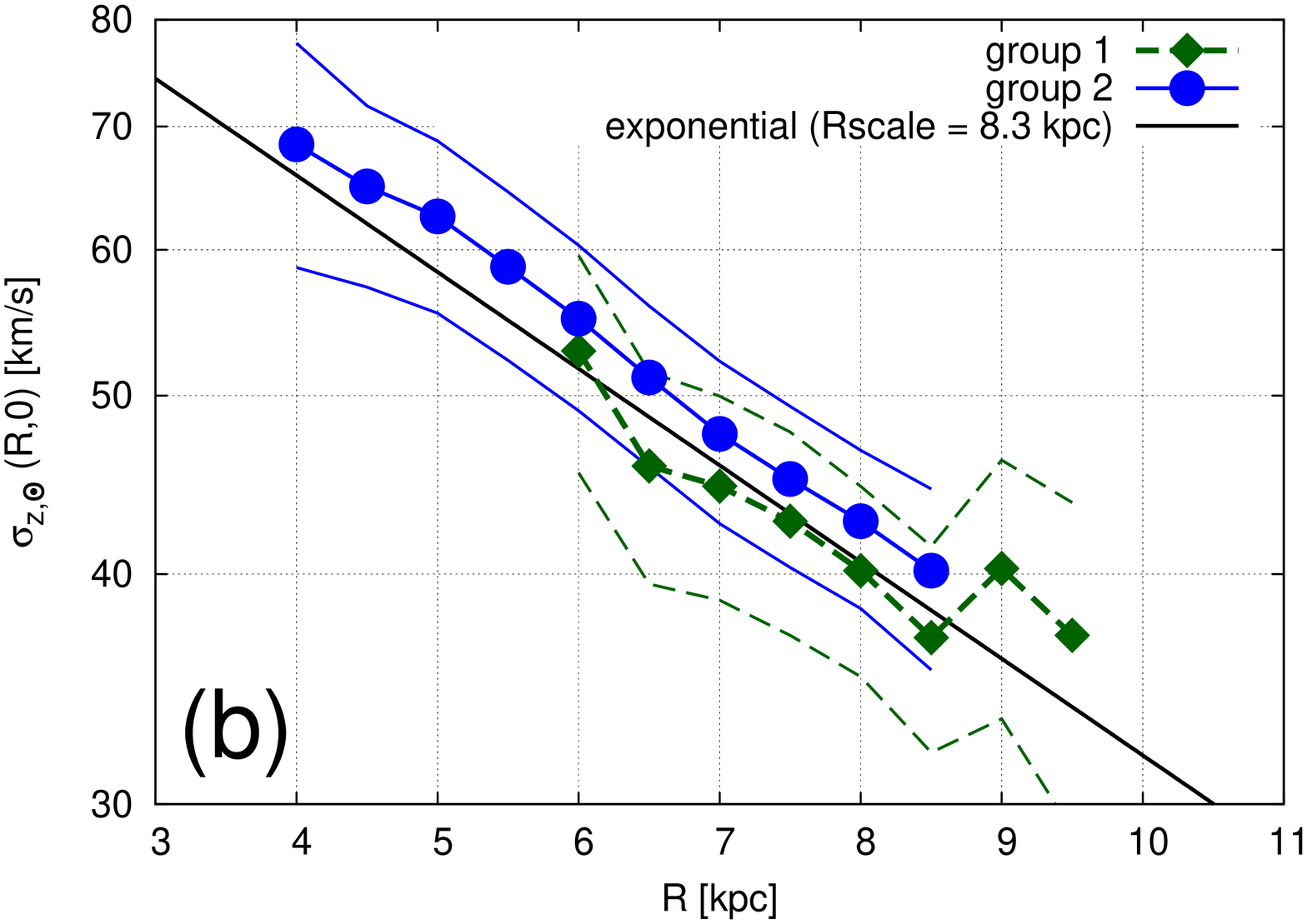} \\
\end{center}
\caption{
(a) 
The distribution of stars in group 1 ($5.5<R_{\rm peri}/{\rm kpc}<6.5$) 
and group 2 ($5.5<(1/2)(R_{\rm peri}+R_{\rm apo})/{\rm kpc}<6.5$) 
in the $(v_\phi, |v_R|)$ plane 
when they pass through $(R,z)=(6.5, 0)$ kpc. 
(b) 
The reconstructed profile of $\sigma_{z,\odot}$ for these groups. 
As a reference, we show the exponential profile 
shown in Figure \ref{fig:sigz_data} (black solid line). 
}
\label{fig:group12}
\end{figure}

As discussed in section \ref{subsection:observable_dispersion}, 
$\sigma_{z,\odot}(R,0)$ 
represents the vertical velocity dispersion of stars 
located at $(R,0)$ 
whose velocity is within the observable region in the velocity space (Figure \ref{fig:observable}). 
In order to see the variation of the vertical kinematics within the observable region, 
we create the following groups of stars from our sample: 
(1) those stars with $5.5<R_{\rm peri}/{\rm kpc}<6.5$ ($N=26$ stars); 
and 
(2) those stars with $5.5<(1/2)(R_{\rm peri}+R_{\rm apo})/{\rm kpc}<6.5$ ($N=30$ stars). 
Here, we define $R_{\rm peri}$ and $R_{\rm apo}$ as the minimum and maximum value of 
$R$ for each star's orbit. 
These samples happen to be non-overlapping, 
and the observed vertical velocity dispersion 
for groups 1 and 2 are respectively 
$38.9\pm 5.4\;{\rm km\;s^{-1}}$ and $42.6 \pm 5.5\;{\rm km\;s^{-1}}$. 
All of these stars spend some time at $(R,z)=(6.5, 0)\;{\rm kpc}$, 
so in Figure \ref{fig:group12}(a), 
we show the distribution of $|v_R|$ and $v_\phi$ 
when they pass through this location. 
The stars in groups 1 and 2 
are distributed in distinct areas in this velocity space, 
reflecting the different orbital eccentricities. 
In Figure \ref{fig:group12}(b), 
we show the vertical velocity dispersion $\sigma_{z,\odot}(R,0)$ of these groups, 
calculated by using $\{w_i\}$ derived in section \ref{section:analysis}. 
We see that the radial profiles of $\sigma_{z,\odot}(R,0)$ 
for these groups are quite similar to each other at $6 \leq R/{\rm kpc} \leq 8.5$. 
This result suggests that  
the vertical kinematics of the stars in the observable region 
is more or less homogeneous at least at $6 \leq R/{\rm kpc} \leq 8.5$.

\section{Discussion} \label{section:discussion}

\subsection{Dependency of the vertical kinematics on radial and azimuthal velocities}

As seen in Figures \ref{fig:velocity_distribution}(a), \ref{fig:velocity_distribution}(b), and \ref{fig:separability}, 
the vertical velocity dispersion of the thick disc
does not have a strong dependency on the in-plane velocities ($v_R$ and $v_\phi$) 
in the Solar neighbourhood. 
In order to have some physical insight into this finding, 
let us suppose, 
as a working hypothesis, 
that this apparent homogeneity of vertical velocity distribution across the $(v_R, v_\phi)$-space 
is the case not only near the Sun but also outside the Solar vicinity. 
Then it is implied that 
the distribution of the vertical action $J_z$ 
is not strongly dependent on the radial action ($J_R$) or azimuthal action ($J_\phi$). 
It follows that there is a typical value of vertical action $J_{z,{\rm typical}}$ 
that characterises the vertical motions of the thick-disc stars. 
On the other hand, 
the maximum height $z_{\rm max}$ above the disc plane 
that a given thick-disc star can reach at Galactocentric radius $R$ 
is determined mainly by its vertical action $J_z$. 
In realistic potential models of the Milky Way, 
$z_{\rm max} \simeq z_{\rm max}(J_z, R)$ is an increasing function of $R$,  
since the gravitational pull towards the disc plane is weaker at larger $R$. 
Thus, 
if the distribution of $J_z$ of the thick-disc stars 
is nearly independent of $(J_R, J_\phi)$, 
then the thick disc needs to flare 
and the scale height of the thick disc 
roughly follows the radial behaviour of $z_{\rm max}(J_{z,{\rm typical}}, R)$.

Then, under what circumstances does 
the distribution of $J_z$ become (approximately) independent of $(J_R, J_\phi)$?
Here we consider such an example. 
Let us suppose that 
the thick disc was formed through radial migration 
\citep{Sellwood2002, Roskar2008, Schonrich2009a, Schonrich2009b, Loebman2011}. 
Some numerical simulations suggest that 
the radial migration conserves the stellar vertical action $J_z$ 
as the star changes its in-plane motion \citep{Solway2012}. 
Thus, radial migration 
redistributes the stars in $(J_R, J_\phi)$-space while keeping $J_z$ unchanged. 
This redistribution in the action space 
is expected to erase the initial dependency of $J_z$-distribution on $(J_R, J_\phi)$ \citep{Minchev2012}, 
unless the initial dependency is too strong to erase. 
If we take into account that the thick-disc stars are generally very old 
($\gtrsim 8$ Gyr; \citealt{Haywood2013}), 
it may be likely 
that the radial migration ends up with 
a homogeneous distribution of $J_z$ as a function of $J_R$ or $J_\phi$. 
It is of interest to note that 
some numerical simulations predict that 
the thick disc flares 
if it was formed though 
radial migration \citep{Loebman2011, Minchev2012, Roskar2013}.

\begin{figure}
\begin{center}
	\includegraphics[angle=-90,width=0.95\columnwidth]{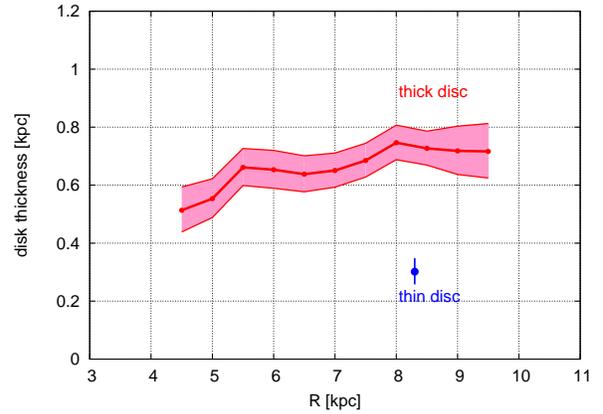} \\
\end{center}
\caption{
The kinematically-estimated thickness of the thick disc as a function of $R$ 
by assuming $\sigma_z(R,0) = \sigma_{z,\odot}(R,0)$. 
Also shown is the kinematically-estimated thickness of the local thin disc 
which is calculated by assuming $\sigma_{z}^{\rm thin} = 20 \pm 2.5\;{\rm km\;s^{-1}}$. 
}
\label{fig:H_data}
\end{figure}

If $\sigma_z$ is really independent of $(v_R, v_\phi)$ anywhere on the disc plane, 
we have $\sigma_z(R,0) \simeq \sigma_{z,\odot}(R,0)$. 
In this case 
we can estimate the thickness $H(R)$ of the thick disc at Galactocentric radius $R$
by using our reconstructed profile of $\sigma_{z,\odot}$. 
As an example, 
we define the thickness $H$ by 
\begin{equation} \label{eq:def_H}
\Phi(R, H(R)) = \Phi(R, 0) + \frac{1}{2} \sigma_z^2 (R, 0),  
\end{equation}
where $\Phi(R,z)$ is the Galactic potential. 
By plugging $\sigma_{z,\odot}$ into $\sigma_z$ in equation (\ref{eq:def_H}), 
we obtain $H(R)$ shown in Figure \ref{fig:H_data}. 
Although the $H(R)$ profile mildly depends on the adopted potential, 
we have confirmed that $H$ increases as a function of $R$ 
independent of the potential parameters 
as long as they are within the range 
explored in section \ref{subsection:result_potential_dependece}.

\subsection {How can we keep the scale height constant?} \label{subsection:constant_H}

As we have discussed in the previous subsection, 
the radial dependence of the thickness of the thick disc 
is related to 
the dependency of the vertical kinematics on the in-plane motions.

By taking into account some observational evidence 
of the thick-disc component in external galaxies with a constant scale height 
\citep{vanderKruit1981, Jensen1982, Comeron2011}, 
let us consider what we would need to keep the scale height $H$ constant. 
Here we concentrate our discussion 
on the situation at a certain radius $R=R_1$ in the inner disc ($R_1<R_0$). 
In order to satisfy $H(R_1)=H(R_0)$, 
we require $\sigma_z(R_1,0)>\sigma_{z,\odot}(R_1,0)$ 
so that 
the vertical velocity distribution is inhomogeneous 
in the $(v_R, v_\phi)$ space. 
Since $\sigma_{z,\odot}$ represents the vertical velocity distribution of 
observable stars (those stars whose velocity is inside the observable region) 
and $\sigma_z$ represents the vertical velocity distribution of all the stars, 
this requirement implies the following situation: 
In the inner disc, 
those thick-disc stars with large $J_\phi$ (large guiding centre radii) 
need to have smaller vertical action 
than those thick-disc stars with small $J_\phi$ and $J_R$.\footnote{
In order for a given star at $R=R_1<R_0$ with small $J_\phi$ 
to be unobservable, 
it has to have small $|v_R|$ (see Figure \ref{fig:observable}) 
and thus small $J_R$.} 
This situation may be attained if 
there are enough 
low-eccentricity, shell-like orbits with large vertical motions in the inner disc.

\section{Conclusions} \label{section:conclusion}

We have investigated an orbit-based method to 
reconstruct the observable distribution function $f_\odot$, 
which describes the distribution of action 
within the portion of action space covered by a local disc sample. 
By applying our method 
to 127 chemically-selected thick-disc stars in the Solar neighbourhood, 
we found that 
the vertical velocity dispersion $\sigma_{z,\odot}$ 
that corresponds to $f_\odot$ 
decreases nearly exponentially as a function of $R$ 
with a scale length of
$R_{\rm s}=8.3\pm1.1 {\rm (rand.)} \pm 1.6 {\rm (sys.)}$ kpc.

We also found that the vertical velocity dispersion $\sigma_z$ of the local thick-disc stars 
shows only weak dependency on the in-plane motions $(v_R, v_\phi)$. 
This apparent homogeneity of $\sigma_z$ in $(v_R, v_\phi)$-space 
may be naturally explained 
if the distribution of the vertical action $J_z$ 
is only weakly dependent on the radial action $J_R$ 
and the azimuthal angular momentum $J_\phi$. 
One possible mechanism to produce such action distribution 
is radial migration, 
where the thick-disc stars are re-distributed in the $(J_R, J_\phi)$-space 
while keeping their $J_z$ unchanged \citep{Minchev2012}.

\section*{Acknowledgments}
KH is supported by Japan Society for the Promotion of Science (JSPS) through 
a Postdoctoral Fellowship for Research Abroad. 
KH thanks Yuzuru Yoshii, Jason Sanders and Shigeki Inoue 
for useful feedback on the original manuscript, 
and thanks Jo Bovy for making \galpy\ publicly available. 
Also, the authors thank the referees for their fruitful comments.

\appendix

\section{Selection function of the survey}

In our analyses, we followed the approach of \cite{SLZ1990} 
to relate the observed kinematic data to the underlying distribution function. 
We note here that our approach owes much to the fact that 
our sample stars are located at the immediate Solar neighbourhood ($d \lesssim 100\;{\rm pc}$).

Let us consider a sample of stars 
taken from an imaginary magnitude-limited survey 
in which all the stars with apparent magnitude 
brighter than a certain value are observed. 
In this case, 
those sample stars at large distances are high-luminosity stars, 
while the sample covers a wider luminosity range for nearby stars. 
Thus, such a sample puts more weight on nearby stars. 
If the kinematical properties are highly inhomogeneous 
within the surveyed region, 
this sample is kinematically biased. 
In general, if the surveyed region is large, 
the survey selection function has to be taken into account 
in order to estimate the kinematic properties of the system (McMillan \& Binney 2012). 
This means that our methodology is not applicable to 
a sample that covers a large volume 
unless the completeness is taken into account, 
since our method implicitly requires that the sample stars are fair representatives of the stellar population. 
However, since our sample is distributed in a small region ($d \lesssim 100$ pc), 
it is safe to assume that the kinematical properties of the thick-disc stars are homogeneous 
inside this region (the thick disc has a scale height of $\sim 1$ kpc and a scale length of $2$-$3$ kpc). 
Therefore, the effect from the (poorly-known) 
survey selection function is relatively small in our sample. 

\label{lastpage}

\end{document}